\documentclass[10pt]{article}
\def\GeV{\,\mathrm{GeV}}

\def\lesssim{{\
\lower-1.2pt\vbox{\hbox{\rlap{$<$}\lower5pt\vbox{\hbox{$\sim$}}}}\ }} 
\def\gtrsim{{\
\lower-1.2pt\vbox{\hbox{\rlap{$>$}\lower5pt\vbox{\hbox{$\sim$}}}}\ }}

\begin{document}


\begin{flushright}
    HIP-2003-29/TH\\
    hep-ph/0305134\\
    May 13, 2003
\end{flushright}

\vspace{5mm}

\begin{center}
{\Large\bf MSSM curvaton in the gauge-mediated SUSY breaking}\\
\vspace{5mm} 
{\large S.~Kasuya$^a$, M.~Kawasaki$^b$, and Fuminobu Takahashi$^b$}\\
\vspace{5mm}
{\it 
$^a$ Helsinki Institute of Physics, P. O. Box 64,
     FIN-00014, University of Helsinki, Finland.\\
$^b$ Research Center for the Early Universe, 
     University of Tokyo, Tokyo 113-0033, Japan.\\
}
\end{center}

\begin{abstract}
We study the curvaton scenario using the MSSM flat directions in the
gauge-mediated SUSY breaking model. We find that the fluctuations in
the both radial and phase directions can be responsible for the
density perturbations in the universe through the curvaton mechanism. 
Although it has been considered difficult to have a successful
curvaton scenario with the use of those flat directions, it is
overcome by taking account of the finite temperature effects, which
induce a negative thermal logarithmic term in the effective potential
of the flat direction.
\end{abstract}

\setcounter{footnote}{1}
\renewcommand{\thefootnote}{\fnsymbol{footnote}}

\section{Introduction}
The curvaton is a scalar field whose fluctuations are
isocurvature-like during inflation, but later converted to an
adiabatic counterpart to explain the structures of the universe 
\cite{curvaton,EKM,EJKM}. The amplitude of density fluctuations of the
curvaton during inflation is estimated as $H_*/\phi_*$, provided that
the curvature scale is small enough, {\it i.e.}, 
$V''(\phi_*) \ll H_*^2$, where $H_*$ and $\phi_*$ denote the Hubble
parameter and the amplitude of the curvaton, respectively. The
subscript `$*$' means that the variable is evaluated when the
cosmological scale is departing the horizon during inflation. 

Among realistic particle theories, the minimal supersymmetric standard
model (MSSM) is the most attractive one, in which there are many
scalar fields called flat directions. The potential of flat directions
vanishes in the supersymmetry (SUSY) exact limit, but it is lifted by
the SUSY breaking effect and some nonrenormalizable operators 
($W \sim \phi^n$, $n=4 - 9$) \cite{DiRaTh}. The flat direction (FD)
field could be a good candidate of a curvaton. The earlier studies
revealed that the curvaton scenario is usually difficult to achieve
along the context of the gravity-mediated SUSY breaking models
\cite{EKM,EJKM}. This is because thermal effects, especially
corrections to the potential, lead to the fact that the energy density
of the FD field will not dominate the universe before it decays. Even
if thermal effects are suppressed as in the case of the hidden
radiation \cite{EKM,EJKM}, the energy density of the FD field does not
dominate the universe except for the $n=9$ direction, and the
amplitude of the fluctuations generated during inflation usually damps
considerably in the course of the evolution after inflation. The
crucial point is that the oscillation of the FD field starts rather
early so that it decays well before it dominates the energy of the
universe.

One may wonder if the oscillation could be `delayed' by some
mechanism. This is exactly achieved in the context of the
gauge-mediated SUSY breaking models. For many flat directions, 
the sign of the two-loop thermal correction to the potential is
negative, which traps the field at a large amplitude until very late
epoch \cite{KKT}. The energy density of the FD field dominates the
universe soon after the FD field oscillation starts. The field
naturally deforms into Q balls, which act as a protector from thermal
scatterings, and have relatively long lifetime. 
In this way, it is possible to have
successful curvaton models with the use of the MSSM FD in the
gauge-mediated SUSY breaking models. We will show how it is actually
realized below. Notice that the key to the problem is the trap due to
this two-loop thermal correction to the potential. If it were not
there (or if it were ineffective), the energy density of the Q balls
would not be able to dominate before the big bang nucleosynthesis
(BBN) and/or the decay would occur before its domination.

\section{Flat direction}
The scalar potential vanishes along the flat direction 
$\Phi=\phi e^{i \theta}/\sqrt{2}$, and it is only lifted by the
gauge-mediated SUSY breaking effect and some nonrenormalizable
operators. It can be written as \cite{KuSh}
\begin{equation}
    \label{pot}
    V(\Phi)=M_F^4\log\left(1+\frac{|\Phi|^2}{M_S^2}\right) 
    +\lambda^2\frac{|\Phi|^{2(n-1)}}{M_P^{2(n-3)}},
\end{equation}
where $M_S$ is the messenger scale, $M_F=(m_{\phi}M_S)^{1/2}$ is the
SUSY breaking scale, $m_{\phi} \sim$ TeV is the mass scale of squarks,
and $M_P=2.4\times 10^{18}$ GeV is the Planck mass.

For the curvaton mechanism to work, the mass of the FD field during
inflation should be negligible compared with the Hubble parameter. It
can be achieved in, say, supersymmetric  D-term inflation models
\cite{KoMa} or the no-scale type inflation~\cite{PLB355-71}, 
but here we just assume that there is no Hubble-induced
mass term during inflation. Then the FD field slow rolls on the
nonrenormalizable potential $V_{NR}$, obeying the equation of motion
\begin{equation}
    3H\dot{\phi}+V'_{NR}(\phi) \simeq 0.
\end{equation}
After inflation, inflaton oscillates around the minimum of its
potential, and the universe is dominated by the inflaton oscillation
energy, which behaves as nonrelativistic matter. During this stage,
inflaton gradually decays into light degrees of freedom, forming
dilute radiation. Although the energy density of this radiation is
very small compared with the total energy density, it actually affects
the potential of the FD field. There are thermal effects at both
one-loop and two-loop order. The former one is a
thermal mass term which is effective for relatively small amplitude of
the flat direction : $\phi \lesssim f^{-1} T$, where $f$ is gauge or
Yukawa coupling constant. The latter one is a two-loop thermal
logarithmic potential that dominates over the thermal mass term for
larger $\phi$. Hereafter we concentrate on the two-loop thermal
effects, which can be written as \cite{AnDi,FuHaYa,KKT} 
\begin{equation}
    V_T = c_T f^4 T^4 \log \left(\frac{|\Phi|^2}{T^2}\right),
\end{equation}
where $c_T$ is a constant of order unity. We expect $f \sim 0.1$,
since the contribution of the gauge interactions dominate over that of
Yukawa interactions. We found that the sign of $c_T$ is negative for
most of the flat directions, so we assume this is the case, setting
$c_T=-1$.  For the sake of completeness, we will discuss the positive
case in App. A, in which there is no successful curvaton scenario.

Because of this negative thermal logarithmic potential, the FD field
will soon be trapped at the amplitude
\begin{equation}
    \phi_m\sim \left(\frac{f^2T^2 M_P^{n-3}}{\lambda}
    \right)^{1\over n-1}.
\end{equation}
It is not released until the zero temperature potential overcomes the 
thermal counterpart, $fT<M_F$. Notice that the Hubble parameter
becomes much smaller than the curvature of the potential at that
time. 

After the oscillation of the FD field commence, it feels spatial
instabilities, and deforms into Q balls right after the oscillation
begins \cite{KuSh,KK1,KK2,KK3,KKT}. The oscillation starts at the
amplitude 
\begin{equation}
    \label{phi-osc}
    \phi_{osc} \sim \left(\frac{M_F^2 M_P^{n-3}}{\lambda}
    \right)^{1\over n-1},
\end{equation}
so that the charge of the produced Q ball becomes
\begin{equation}
    Q \sim \beta \left(\frac{\phi_{osc}}{M_F}\right)^4
    \sim \beta \left(\frac{M_P^{n-3}}{\lambda M_F^{n-3}}
    \right)^{4\over n-1}.
\end{equation}
where $\beta \lesssim 0.1$. \footnote{
Since the Hubble parameter is much smaller than the curvature of the
potential at the onset of the oscillation, we expect 
$\beta \lesssim 0.1$ instead of $\beta \sim 10^{-4}$ \cite{KK3}.}
Here and hereafter we assume that $\phi_{osc} \gtrsim M_S \sim 
M_F^2/m_{\phi}$ is satisfied.

If the universe has already become radiation-dominated when 
$T \sim f^{-1}M_F$, the energy density of the Q balls 
the FD condensate is only $f^4$ times smaller than that of the
radiation. Hence  the energy of the Q balls will dominate the universe
soon because it decreases as $\propto a^{-3} \propto T^3$, while the
radiation density decreases as $\propto a^{-4} \propto T^4$. Thus,
the energy density of the Q balls becomes equal to that of radiation
when   
\begin{equation}
    \label{teq1}
    T=T_{eq} \sim f^3 M_F.
\end{equation}
On the other hand, if the FD field starts its oscillation during the
inflaton oscillation dominated (IOD) universe, the energy density of
the Q balls evolves as $a^{-3}$, while the radiation decreases as
$a^{-3/2}$. The energy density of the Q balls will not dominate the
universe until
\begin{equation}
    \label{teq2}
    T=T_{eq} \sim f^3 \left(\frac{T_{RH}}{f^{-1}M_F}\right)^5 M_F.
\end{equation}
Notice that $T_{RH} \lesssim f^{-1} M_F$ in this case. For the curvaton
scenario to work, the temperature $T_{eq}$ must be higher than the
decay temperature $T_d$ (see Eqs.(\ref{Td}) and (\ref{Td2})). 

The Q balls can decay if the mass per unit charge is larger than the
mass of the decay particle $m_d$. This condition is expressed as
\begin{eqnarray}
    \label{decay-cond}
    M_F Q^{-\frac{1}{4}} > m_d.
\end{eqnarray}
In the case of the Q ball with $B \ne 0$, nucleons must be in the
decay particles, so that $m_d = m_N \sim 1$~GeV, and the condition
becomes 
\begin{eqnarray}
    M_F &>& \left( \beta^{\frac{n-1}{4}} \lambda^{-1} m_N^{n-1} 
      M_P^{n-3} \right)^{\frac{1}{2(n-2)}} ,\nonumber \\
    &\sim& 
    \left \{
      \begin{array}{ll}
          3 \times 10^4~\GeV & ~~{\rm for~~~}n = 4,\\
          9 \times 10^5~\GeV & ~~{\rm for~~~}n = 5,\\
          5 \times 10^6~\GeV & ~~{\rm for~~~}n = 6,
      \end{array}
    \right.
\end{eqnarray}
where we set $\beta = 0.1$ and $\lambda =1$. If the Q ball has no
baryon number $B=0$ but its constituent includes squarks, the decay
particles are pions; $m_N$ should replaced by $m_{\pi} \sim 0.1$~GeV.
If the Q ball is non-baryonic, it can decay into neutrinos, and
essentially there is no condition like Eq.(\ref{decay-cond}).

The Q ball decays by loosing its charge through the surface with the
rate \cite{Cohen}
\begin{equation}
    \label{Q-decay}
    \Gamma_Q \sim \frac{M_F}{48\pi Q^{5\over 4}}, 
\end{equation}
so that the decay temperature is given by
\begin{equation}
    \label{Td}
    T_d \sim \left( \Gamma_Q M_P \right)^{1\over 2}
        \sim \frac{\lambda^{\frac{5}{2(n-1)}}}{\sqrt{48\pi}}
         \beta^{-\frac{5}{8}}
             \left(\frac{M_F}{M_P}\right)^{3n-8 \over n-1} M_P.
\end{equation}
Imposing $T_d \gtrsim$ MeV so that the BBN is successful, we must have
$M_F$ larger than $10^3$, $3\times 10^6$ and $10^8$ GeV 
for $n=4$, 5, and 6, respectively. For larger $n$, Q balls cannot
decay before the BBN time. Notice that there is another constraint
from the gauge-mediated SUSY breaking models; 
$M_F \lesssim \sqrt{g m_{3/2}M_P}$, {\it i.e.}, 
$M_F \lesssim 10^8$ GeV for $m_{3/2}\lesssim$GeV, where $g$ is a
gauge coupling. For $n=4$, more stringent bound appears as $M_F
\lesssim 10^7$ GeV, which comes from the condition 
$\phi_{osc} \gtrsim M_S \sim M_F^2/m_{\phi}$.  

On the other hand, if the Q-ball charge is small enough, such that 
$Q<Q_c \simeq 10^{16}(M_F/10^7{\rm GeV})^{-4/11}$ \cite{KK3}, Q-balls
decay through thermal effects. In this case, the decay temperature can
be written as \cite{KK3}
\begin{equation}
    \label{Td2}
    T_d \sim 10\frac{M_p}{Q}.
\end{equation}
Notice that this decay process is realized only for $n=4$ with
$M_F \gtrsim 3 \times 10^4$ GeV.

\section{Dynamics of fluctuations}
\subsection{Fluctuations in the radial direction}
The second important aspect of the curvaton mechanism is to produce 
enough amount of the fluctuations during inflation, {\it i.e.}, 
$\delta\equiv H_*/\phi_* \gtrsim 10^{-5}$. In general, the position of
FD field during inflation is constrained as $V'' \lesssim H^2$,
leading to $\phi_* \lesssim \phi_{sr}\equiv
(H M_P^{n-3}/\lambda)^{1/(n-2)}$. Hereafter we assume that $\phi_*$ is
of the order of $\phi_{sr}$ for a definite
discussion~\cite{Kawasaki:2001in}. Thus
\begin{equation}
    \frac{H_*}{\phi_*} \sim \lambda 
    \left(\frac{\phi_*}{M_P}\right)^{n-3},
\end{equation}
leading to
\begin{eqnarray}
    \phi_* & \sim & 
    \left(\frac{\delta}{\lambda}\right)^{1 \over n-3} M_P, \\
    H_* & \sim & \lambda^{-\frac{1}{n-3}} \delta^{n-2 \over n-3} M_P.
\end{eqnarray}
As we will see shortly, however, the amplitude of the
fluctuations will damp in the course of the evolution after
inflation. In general, the equations of motion of the homogeneous and
fluctuation modes are given by
\begin{eqnarray}
    \label{phi-ho-sr}
    & & \ddot{\phi} + 3H\dot{\phi} + V'(\phi) = 0, \\
    \label{phi-fl-sr}
    & & \delta\ddot{\phi} + 3H\delta\dot{\phi} 
    + V''(\phi) \delta\phi= 0, 
\end{eqnarray}
where the wave number $k\rightarrow 0$ for the superhorizon mode is
assumed for the fluctuation. After inflation the homogeneous mode of
the FD field still (marginally) slow rolls in the nonrenormalizable
potential during IOD, and the damping factor is estimated as
\begin{equation}
    \frac{\delta\phi}{\phi} \sim 
    \left(\frac{\delta\phi}{\phi}\right)_i
    \left(\frac{\phi}{\phi_i}\right)^{n-4 \over 2}\,,
\end{equation}
where the subscript $i$ denotes the initial values. This slow-roll
regime will end when the field is trapped by the negative thermal
logarithmic potential. Since the curvature at the minimum is
\begin{equation}
    V''(\phi_m)\sim \left[\frac{\lambda(fT)^{2(n-2)}}{M_P^{n-3}}
      \right]^{2 \over n-1},
\end{equation}
and $T \sim (T_{RH}^2H M_P)^{1/4}$ during inflaton oscillation
dominated universe, the trap starts when
\begin{equation}
    H=H_{tr} \sim \left( \frac{\lambda^2 f^{4(n-2)} T_{RH}^{2(n-2)}}
      {M_P^{n-4}} \right)^{1 \over n}.
\end{equation}
The amplitude of the field and the temperature at that moment are
\begin{eqnarray}
    \phi_{tr} & \sim & \left(\frac{f^4 T_{RH}^2 M_P^{n-2}}{\lambda}
      \right)^{1 \over n}, \\
    \label{Ttr}  
    T_{tr} & \sim & \left( \lambda^{1\over2} f^{n-2}
      T_{RH}^{n-1} M_P \right)^{1\over n},
\end{eqnarray}
respectively, so that the damping factor during slow roll in the
nonrenormalizable potential is estimated as 
\begin{equation}
    \label{dphi-sr}
    \xi_{\phi}^{(SR)} \equiv
    \frac{\left(\frac{\delta\phi}{\phi}\right)_{tr}} 
         {\left(\frac{\delta\phi}{\phi}\right)_*} \sim
    \left(\frac{\phi_{tr}}{\phi_*}\right)^{n-4 \over 2} \sim
    \left[
    \left(\frac{f^4 T_{RH}^2}{\lambda M_P^2}\right)^{1 \over n}
    \left(\frac{\lambda}{\delta}\right)^{1 \over n-3}
    \right]^{n-4 \over 2}.
\end{equation}
This result is applicable if the amplitude of the field during
inflation, $\phi_*$, is larger than that of the minimum determined by
the balance of nonrenormalizable and negative thermal logarithmic
terms in the potential, namely
\begin{equation}
    T_{RH} \lesssim T_{RH}^C \equiv 
    \lambda^{-\frac{3}{2(n-3)}} f^{-2} \delta^{n \over 2(n-3)} M_P.
\end{equation}
In the opposite case, the field fast rolls in the negative thermal
logarithmic potential, and quickly settles down to the minimum. In
this course, the amplitude of $\delta\phi/\phi$ does not evolve
so much. Typically, it only decreases an order of magnitude at most.

Once the field is trapped at the instantaneous minimum $\phi_m$, it
is not released until $T\lesssim f^{-1}M_F$. During this stage, the
amplitude of fluctuation compared to the homogeneous mode decreases as
\begin{equation}
    \frac{\delta\phi}{\phi} \propto \left\{
      \begin{array}{l}
          H^{\frac{3n-4}{4(n-1)}}, \qquad {\rm (IOD)}, \\
          H^{\frac{n-3}{4(n-1)}}, \qquad {\rm (RD)}. \\
      \end{array}
      \right.
\end{equation}
Thus, the additional damping factor during the trap is estimated as
\begin{eqnarray}
    \xi_{\phi}^{(tr)} \equiv
    \frac{\left(\frac{\delta\phi}{\phi}\right)_{osc}} 
         {\left(\frac{\delta\phi}{\phi}\right)_{tr}} & \sim &
    \left(\frac{H_{RH}}{H_{tr}}\right)^{\frac{3n-4}{4(n-1)}}
    \left(\frac{H_{osc}}{H_{RH}}\right)^{\frac{n-3}{4(n-1)}},
    \nonumber \\ & \sim &
    \lambda^{-\frac{3n-4}{2n(n-1)}}f^{-\frac{7n-16}{2n}}
    T_{RH}^{-\frac{n-8}{2n}}M_F^{\frac{n-3}{2(n-1)}}
    M_P^{-\frac{3n-4}{n(n-1)}},
\end{eqnarray}
where we assumed that the release of the FD field from the trap
by the negative thermal logarithmic potential takes place during
radiation-dominated universe, {\it i.e.}, $T_{RH}>f^{-1}M_F$. Thus the
amplitude of the fluctuations damps by a factor
$\xi_{\phi}^{(SR)}\xi_{\phi}^{(tr)}$ during both the slow-roll and the
trap periods. Since $\xi_{\phi}^{(SR)}$ has dependences on the
reheating temperature as
\begin{equation}
    \xi_{\phi}^{(SR)} \propto \left\{
      \begin{array}{ll}
          T_{RH}^{n-4 \over n} & (T_{RH} < T_{RH}^C), \\[2mm]
          {\rm const.} & (T_{RH} > T_{RH}^C), 
      \end{array}
      \right.
\end{equation}
the total damping factor behaves according to
\begin{equation}
    \xi_{\phi}^{(SR)}\xi_{\phi}^{(tr)} \propto \left\{
      \begin{array}{ll}
          T_{RH}^{1 \over 2} & (T_{RH} < T_{RH}^C), \\[2mm]
          T_{RH}^{\frac{8-n}{2n}} & (T_{RH} > T_{RH}^C).
      \end{array}
      \right.
\end{equation}
At a glance, higher reheating temperature than $T_{RH}^C$ seems to
give less damping factor, but it will be shown shortly that it does
not open parameter space for successful scenario.

To this end, let us estimate the damping factor when
$T_{RH}=T_{RH}^C$. In this case we can set $\xi_{\phi}^{(SR)}\sim 1$.
The amplitude of the fluctuation can be estimated as
\begin{equation}
    \tilde{\delta}\equiv \left.\frac{\delta\phi}{\phi}\right|_{osc} 
    \sim \lambda^{-\frac{3n+1}{4(n-1)(n-3)}} f^{-\frac{5}{2}}
    \left(\frac{M_F}{M_P}\right)^{n-3 \over 2(n-1)}
    \delta^{3n-4 \over 4(n-3)}.
\end{equation}

Now let us look at each case of $n=4,~5$, and 6 when the Q-ball charge
is larger than $Q_c$. For $n=4$, $\tilde{\delta}\sim 10^{-5}$ can be
realized if we take $\lambda \sim 1$, $f \sim 0.03$, 
$M_F \sim 3 \times 10^4$ GeV, and 
$\delta \sim 5 \times 10^{-4}$, for example. In this case, $T_{RH}
\sim T_{RH}^C \sim 10^{15}$ GeV. Since $T_d \sim 30$ MeV while $T_{eq}
\sim$ GeV, the Q balls decay after they dominate the energy of the
universe, so that one of the conditions for the curvaton mechanism to
work is satisfied. However, the gravitinos are overproduced in such high
reheating temperature, and it is necessary to dilute their density by
a factor of $(T_{RH}/T_{3/2}) \sim 10^6$, where  $T_{3/2} \sim 10^9$
GeV for $m_{3/2} \sim$ GeV, is the upper limit of the reheating
temperature in which the cosmological gravitino problem is avoidable
\cite{Moroi}. Although there is some entropy production when Q balls
decay, the dilution factor is only $\sim 30$, and the abundance of
gravitinos is still unacceptably large.\footnote{
Since there is no damping effect during the slow-roll regime in the
case of $n=4$, it seems possible to have lower reheating temperature 
$\sim 10^9$~GeV such that the gravitino problem can be evaded. 
However, the coupling constant $f$ should be unnaturally small 
$\sim 0.003$ in order to realize $T_{eq}<T_d$ in this case.}

We can analyze $n=5$ and 6 cases in similar manners. For $n=5$, 
$\tilde{\delta} \sim 10^{-5}$ can be explained by setting 
$\lambda \sim 1$, $f\sim 0.1$, $M_F\sim 10^8$ GeV, and 
$\delta \sim 2 \times 10^{-4}$. However, the reheating temperature is
very high $\sim 10^{16}$ GeV, which is highly speculative. In
addition, the entropy production due to the decay of Q balls is not
enough to dilute the overproduced gravitinos. For $n=6$, the parameter
set of $\lambda \sim 1$, $f\sim 0.1$, $M_F \sim 10^8$ GeV, and 
$\delta \sim 10^{-4}$ lead to the right amount of the fluctuations, 
{\it i.e.}, $\tilde{\delta} \sim 10^{-5}$, but still very high
reheating temperature such as $\sim 10^{16}$ GeV is needed. 

To sum up, $n=4$ direction could be a curvaton provided that there was
some other entropy production. The cases of $n=5$ and 6 are hopeless, 
since too high reheating temperature such as $\sim 10^{16}$ GeV is
necessary, where the adiabatic fluctuations of the inflaton should
dominate among others. As for the larger reheating temperature than 
$T_{RH}^C$ case, it cannot be realized in curvaton context for $n=5$
and 6, because $T_{RH}$ exceeds $10^{16}$ GeV. For $n=4$, $T_{RH}$
can be raised an order of magnitude at most, but it only makes 3 times
less damping effect, and does not essentially change our result here.

However, the situation changes dramatically in the case of $Q<Q_c$ for
$n=4$ directions, in which the Q ball decays through thermal effects.
The ideal example is $QuQd$ direction \footnote{
The negative thermal logarithmic potential for this direction is shown
in App. B.}
which does not carry any
baryonic or leptonic charge, thus we do not have to care about the
problematic baryonic isocurvature fluctuations. Since there is no
damping effect during the slow roll regime in the case of $n=4$, we
only have to estimate the damping factor during the trap due to the
negative thermal log potential. For $\lambda\simeq 10^{-2}$, 
$f\simeq 0.2$, $M_F=10^7$ GeV, and $T_{RH}\simeq 10^{11}$ GeV, the
charge of the Q ball becomes $\sim 10^{16}$, and the decay temperature
is as high as $\sim 10^3$ GeV, while the energy density of the Q ball
dominates the universe when $T=T_{eq}\simeq 10^5$ GeV. Thus, enough
entropy will be released through the Q-ball decay (dilution factor
$\sim 100$) to evade the gravitino problem. As for the amplitude of
the fluctuation, we can obtain $\tilde{\xi} \sim 10^{-5}$ for 
$\phi_* \sim 10^{-3}M_p$ and $H_* \sim 5 \times 10^{-5} M_p$. Since
the decay temperature is higher than the electroweak scale, the
baryogenesis may take place in the course of the electroweak phase
transition.

Finally we comment on the case that the FD field is released from the
trap during IOD ({\it i.e.}, $T_{RH} < f^{-1}M_F$). In this case it is
impossible to adjust parameters to be $\phi_* \simeq \phi_{tr}$, hence
the damping effects is unavoidable. Also, the later energy 
dominance by the Q balls is difficult to achieve while keeping the
amplitude of the fluctuations larger than $10^{-5}$. Thus, there is no
successful scenario.

\subsection{Fluctuations in the phase direction}
\label{subsec:phase}

The FD field generally has the A-terms in the potential. For the
superpotential $W_{NR}\sim \lambda\Phi^n/M_P^{n-3}$, they are written
as 
\begin{equation}
    V_A \sim \frac{\lambda m_{3/2} \Phi^n}{M_P^{n-3}} + {\rm h.c.},
\end{equation}
where we assumed the vanishing (or negligible) cosmological
constant in the vacuum. As is well known, such an A-term 
appears only if the nonrenormalizable superpotential is
$\phi^n$-type. If it is $\chi \phi^{n-1}$-type, where $\chi$ is any
(super)field other than those consist of the flat direction, no A-term
results. The  $\chi \phi^{n-1}$-type superpotential appears for all
$n=5$ flat directions \cite{GhKoMa}. (Note that this fact does not
mean that there is no A-term in the case of $n=5$; $\phi^{\tilde{n}}$
type superpotential with $\tilde{n}>5$ induces the A-terms, although
they are suppressed by $(\Phi/M_P)^{\tilde{n}-n}$. In fact, we have
found that there is no successful scenario in the case of $n=5$.)
Therefore, we will consider $n=4$ and 6 cases here. Since the
fluctuation of the potential becomes
\begin{equation}
    \Delta V(\theta) \sim \frac{\lambda m_{3/2} \phi_{osc}^n}
    {M_P^{n-3}} \ \sin n\theta \ \delta\theta \big|_{osc},
\end{equation}
at the onset of the FD field oscillations, the density perturbation
is estimated as
\begin{equation}
    \frac{\delta\rho}{\rho} \sim  
    \frac{\Delta V(\theta)}{V(\phi_{osc})} \sim 
    \lambda^{-\frac{1}{n-1}} \ \left(\frac{m_{3/2}}{M_P}\right) \
    \left(\frac{M_F}{M_P}\right)^{-\frac{2(n-2)}{n-1}}
    \sin n\theta \ \delta\theta \big|_{osc},
\end{equation}
where $V(\phi_{osc})\sim M_F^4$ and Eq.(\ref{phi-osc}) are used. 

Now we discuss the evolution of the fluctuations. Since the amplitude
of the adiabatic fluctuations is determined by $\theta$ and 
$\delta\theta$, not the combination like $\delta\theta/\theta$, we
need follow the evolution of  $\theta$ and $\delta\theta$ in the
potential $V(\eta) \sim \lambda m_{3/2}\phi^{n-2} \eta^2/M_P^{n-3}$,
where $\eta=\theta$ or $\delta\theta$. The mass scale of the phase
direction, $m_{\theta}$, remains smaller than the Hubble parameter
until well after the radial direction $\phi$ is trapped at the
instantaneous minimum $\phi_m$, and the amplitudes of $\theta$ and
$\delta\theta$ will not decrease during the slow roll. The damping
takes place only at the very late stage, just before $\phi$ is
released from the trap. This happens when the slow roll condition on
the phase direction is violated, $m_{\theta} > H$, and hence 
\begin{equation}
    \label{Trh}
    T < T_{sr} \equiv \left\{
    \begin{array}{ll}
        \left( \lambda f^{2(n-2)} m_{3/2}^{n-1} T_{RH}^{4(n-1)}
          M_P^{n+1} \right)^{1 \over 2(3n-2)} & {\rm (IOD)}, \\
        \left( \lambda f^{2(n-2)} m_{3/2}^{n-1} M_P^{n+1} 
          \right)^{1 \over 2n} & {\rm (RD)}.
    \end{array}
    \right.
\end{equation}
After the temperature drops down to this value, the amplitude $\eta$
($=\theta$ or $\delta\theta$) will decrease obeying the equation
\begin{equation}
    \ddot{\eta}+3H\dot{\eta}
          +2\frac{\dot{\phi}}{\phi}\dot{\eta} 
          +m_{\theta}^2\eta = 0,
\end{equation}
where we assumed that $\delta\phi$ has died out completely. It is 
easily derived that
\begin{equation}
    \eta \propto \left\{ 
    \begin{array}{ll}
        H^{7n-10 \over 8(n-1)} \propto T^{7n-10 \over 2(n-1)} 
        & {\rm (IOD)}, \\
        H^{2n-5 \over 4(n-1)} \propto T^{2n-5 \over 2(n-1)} 
        & {\rm (RD)}.
    \end{array}
    \right.
\end{equation}

First we consider the case that the trap ends after the reheating,
{\it i.e.}, $T_{RH} > T_{sr}$. Since the damping effect is very mild
during the radiation-dominated era compared to
inflaton-oscillation-domination, one can avoid considerable damping of
$\theta$ and $\delta\theta$ in this case. In fact, we obtain a damping
factor
\begin{equation}
    \xi_{\theta} \equiv 
    \frac{\left(\theta \delta \theta\right)_{osc}}
         {\left(\theta \delta \theta\right)_{*}} 
    \sim \left(\frac{T_{osc}}{T_{sr}}\right)^{2n-5 \over n-1}
    \sim \lambda^{-\frac{2n-5}{2n(n-1)}} f^{-\frac{2(2n-5)}{n}}
    \left(\frac{m_{3/2}}{M_P}\right)^{-\frac{2n-5}{2n}}
    \left(\frac{M_F}{M_P}\right)^{\frac{2n-5}{n-1}},
\end{equation}
considering both $\theta$ and $\delta\theta$. Thus, the amplitude of
the fluctuations can be estimated as
\begin{equation}
    \frac{\delta\rho}{\rho} \sim  
    \lambda^{-\frac{4n-5}{2n(n-1)}} f^{-\frac{2(2n-5)}{n}}
    \left(\frac{m_{3/2}}{M_P}\right)^{5 \over 2n} \
    \left(\frac{M_F}{M_P}\right)^{-\frac{1}{n-1}}
    \theta_* \ \frac{H_*}{\phi_*},
\end{equation}
where $\theta_*$ is the value of $\theta$ during inflation, and 
$\delta\theta \sim H_*/\phi_*$ is used. It is easily seen that
successful curvaton scenario is achieved only for $n=6$. For example,
the parameter set of $\lambda \sim 1$, $f\sim 0.05$, $M_F\sim 10^8$~GeV, 
$m_{3/2}\sim 1$~GeV, $\theta_* \sim 0.3$ and $\delta \sim 10^{-2}$ 
\footnote{
$\delta\equiv H_*/\phi_*$ is a little larger than that value
determined by the condition that the FD field starts slow rolling, but
it is possible for the FD field to have an order of magnitude
smaller amplitude due to its dynamics~\cite{Kawasaki:2001in}.} 
realizes the desired density perturbation,
$\delta\rho/\rho \sim 10^{-5}$, where the damping factor
$\xi_{\theta}$ is $\sim 0.3$. Notice that the disastrous baryonic
isocurvature problem is naturally avoided because the baryon
asymmetry vanishes due to the nonlinear dynamics of the Q-ball
formation, in spite of the fact that any $n=6$ FD field has non-zero  
$B$. This happens when the oscillation of the FD field starts after 
$H < m_{\theta}$ in negative thermal logarithmic potential
\cite{KKT}.

For the parameter set exemplified above, the reheating temperature
is constrained as $T_{RH} > T_{sr} \sim 7 \times 10^9$ GeV.
Such high reheating temperature leads to overproduction of the
gravitino by a factor $\sim (T_{RH}/T_{3/2}) \sim 7$.
However, the Q balls dominate the universe later, and enormous
amount of entropy will be released. Since the domination by the Q
balls begins soon after the oscillation of the FD field starts, the
dilution factor becomes 
$(H_{eq}/H_d)^{1/2} \sim (f^3 M_F/T_d) \sim 2\times 10^5$, where
$T_{eq} \sim 600$ GeV and  $T_d \sim 3$ MeV (see below Eq.(\ref{Td})),
which is enough to dilute the overproduced gravitinos.
Notice that we just require the reheating temperature higher than 
$T_{sr}$(RD). Of course, there is an upper limit in order not to have
gravitino overproduction. Thus,  
$7\times 10^9$ GeV $\lesssim T_{RH} \lesssim 10^{13}$ GeV is
allowed.\footnote{
If the messenger scale $M_S$ is less than the reheating temperature,
the particles in the messenger sector give significant contributions  
to the gravitino production~\cite{Choi}, 
which make the constraint more stringent.
However, in the present case, $M_S \sim M_F^2/m_{\phi} \sim 
10^{13}$~GeV $ > T_{RH}$, and hence the messenger contributions 
are neglected. }

Now let us consider the cases that the trap ends during IOD. Then the
damping factor becomes 
\begin{equation}
    \xi_{\theta} \sim 
    \left(\frac{T_{osc}}{T_{sr}}\right)^{7n-10 \over n-1}
    \sim \left[ \lambda^{-1} f^{-4(n-1)}
    \left(\frac{T_{RH}}{f^{-1}M_F}\right)^{-4(n-1)}
    \frac{M_F^{2n}}{m_{3/2}^{n-1} M_P^{n+1}}
    \right]^{7n-10 \over 2(n-1)(3n-2)}.
\end{equation}
The damping is tremendous for $n=4$, since 
$\xi_{\theta} \lesssim 10^{-10}$ even for $M_F \sim 3\times 10^4$ GeV,
$f\sim 10^{-2}$, and $T_{RH}/(f^{-1}M_F) \lesssim 1$. For $n=6$, one
could obtain the right amount of the fluctuations. It is achieved if
we set the parameters as, for example, $f \sim 0.1$, $M_F\sim 10^8$
GeV, $T_{RH}\sim 5 \times 10^8$ GeV, and $\delta \sim 10^{-2}$. In
this case, $T_{eq} \sim 3 \times 10^3$ GeV and $T_d \sim 1$ MeV for
$\beta \sim 0.1$. This situation is very similar to the above $n=6$
case when the trap ends during RD, since the baryon asymmetry is not
created because of the nonlinear dynamics of Q-ball formation. 
However, it is different in that there is no gravitino overproduction
for relatively low reheating temperature. An example of $n=6$ is $LLe$
direction, and the derivation of the thermal potential is given in the
App. B.

Finally we must mention that this mechanism of generating the density
perturbation using the phase direction is independent of any damping
effects on the radial direction, so that a negative Hubble-induced
mass term can appear during and after inflation. Of course, one must
avoid Hubble-induced A-terms in order that $\theta$ and $\delta\theta$
should not decrease considerably. They appear only if there is
three-point interaction in nonrenormalizable K\"ahler potential, and
these terms do not exist if the inflaton carries some nontrivial
charge.

\section{Conclusion}
We have investigated the possibility of curvaton scenario with the use
of the MSSM flat directions in the gauge-mediated SUSY breaking
models, and shown that the both radial and phase directions of the
flat directions can act as a curvaton. The later energy domination
(before the decay) is usually very difficult to achieve in the
realistic particle physics. In addition, thermal effects such as
corrections to the potential and scatterings leading to early decay
(or evaporation) of the field make the scenario more difficult to be
constructed. In our scenario, both of them are overcome by the 
{\it thermal} effects and Q balls. The FD field dominates the universe
soon after the release from the rather long trap by the negative 
{\it thermal} logarithmic potential. Thermal scatterings do not
affects so much, since the FD field deforms into Q balls soon after
the oscillation starts. The Q balls in the gauge-mediated SUSY
breaking have a long lifetime, and survive from too early evaporation
due to thermal scatterings, if the charge is large enough. On the
other hand, they can decay before BBN (and even electroweak phase
transition) for small enough charge. We have utilized the intermediate
region suitable for the curvaton scenario.  

We have shown that the radial component of the flat direction can play
a role of curvaton in the case of $n=4$. For larger $n$, 
$\delta\phi/\phi$ damps considerably in the higher power potential,
and we have found no successful scenario. In addition, the
fluctuations in the phase direction do not damp so much and can be
responsible for the right amount of the adiabatic perturbation of
order $\sim 10^{-5}$ in the case of $n=6$, provided that there are no
Hubble-induced A-terms. Although such terms can appear due to
nonrenormalizable three-point interactions $\sim I\Phi^\dagger \Phi$
between inflaton $I$ and FD field in the K\"ahler potential, it can be
easily forbidden with use of some symmetry principles.

Furthermore, there is another problem associated with the baryonic
isocurvature perturbation. Except for three $n=4$ directions, 
flat directions possess non-zero $B$ (and $B-L$), which usually leads
to too much baryonic isocurvature perturbation. In the case of the
phase directions as a curvaton, the vanishing baryon asymmetry is
achieved through non-perturbative dynamics of producing Q balls
\cite{KKT}, in which the negative thermal logarithmic potential plays
the crucial role. However, baryogenesis may be still problematic in
this case. In order to avoid large baryonic isocurvature fluctuations,
the baryogenesis must take place after the Q-ball dominates the
Universe. In our scenario the Q-ball dominated Universe begins at the
temperature $T_{eq}\sim 6 \times10^2$ or $3\times 10^3$~GeV, depending
on the thermal history (see Sec.\ref{subsec:phase}). Moreover, in
order for baryon number to survive dilution by the entropy production
due to the Q-ball decay, large baryon-to-entropy ratio should be
generated. Unfortunately, at the moment we do not know such effective
baryogenesis mechanism which works at low energy scales. 

On the other hand, the $n=4$ FD fields such as $QuLe$ and $QuQd$
possess no baryon number, so that there is no baryonic isocurvature
perturbations from the first place in the case of the radial direction
as curvaton. Also, the decay temperature can be higher than the
electroweak scale, therefore the baryogenesis may occur through the
electroweak phase transition. We can thus conclude that these
directions are most promising candidate of a curvaton.

\appendix

\setcounter{equation}{0}
\makeatletter
  \renewcommand{\theequation}{%
        \thesection.\arabic{equation}}
  \@addtoreset{equation}{section}
\makeatother

\section{Positive thermal logarithmic potential}
Here we comment on the case in which the two-loop thermal correction
to the potential has a positive sign, and show that there is no
successful scenario. We have found that the positive potential seems
to appear only for some $n=4$ directions, so we restrict our
discussion only to $n=4$, although some formulae will be given in
general $n$. The FD field starts its oscillation earlier in the
positive logarithmic potential, and the charge of the produced Q ball
is rather small. It is estimated as
\begin{equation}
    Q \sim \beta \left(\frac{\phi_{osc}}{f T_{osc}}\right)^4 
      \sim \left\{
        \begin{array}{ll}
            \beta \lambda^{-\frac{6}{n}}\left(\frac{M_P}{f^2T_{RH}}
              \right)^{4(n-3) \over n} & ({\rm IOD}), \\[2mm]
            \beta \lambda^{-2} f^{-4(n-3)} & ({\rm RD}),
        \end{array}
        \right.
\end{equation}
where the field oscillation starts during IOD and RD at the
temperature and amplitude
\begin{eqnarray}
    T_{osc} & \sim & \left\{
      \begin{array}{ll}
          \left( \lambda^{1\over 2}f^{n-2}T_{RH}^{n-1} M_P 
            \right)^{1\over n}, & ({\rm IOD}), \\[2mm]
         \lambda^{\frac{1}{2}} f^{n-2} M_P, & ({\rm RD}), 
      \end{array}
      \right. \\
    \phi_{osc} & \sim & \left\{
      \begin{array}{ll}
          \left( \lambda^{-1}f^4T_{RH}^2 M_P^{n-2}\right)^{1\over n}, 
          & ({\rm IOD}), \\[2mm]
          f^2 M_P, & ({\rm RD}),
      \end{array}
      \right.
\end{eqnarray}
respectively. The decay process is determined by the evaporation
(and/or diffusion) for the Q ball with small charge, while the Q ball
decays at the rate (\ref{Q-decay}) when the charge is large enough to
survive from the evaporation. Then the decay temperature is
\begin{equation}
    T_d \sim \left\{
    \begin{array}{ll}
        10 \frac{M_p}{Q}, & (Q < Q_c), \\
        \left( \frac{M_F M_P}{48\pi Q^{5\over 4}}\right)^{1\over 2},
        & (Q > Q_c), 
    \end{array}
    \right.
\end{equation}
where the survival condition is 
$Q > Q_c \sim 10^{16}(M_F/10^7 \GeV)^{-4/11}$ \cite{KK3}. As the
energy density of the Q ball equals to that of radiation when
\begin{equation}
    T_{eq} \sim \left\{
      \begin{array}{ll}
          \lambda^{-\frac{5}{2n}} f^{-\frac{2(n-5)}{n}}
          \left(\frac{T_{RH}}{M_P}\right)^{5\over n} M_F, 
           & ({\rm IOD}), \\
           f^{3} M_F,  & ({\rm RD}), \\
      \end{array}
      \right.
\end{equation}
it is easily seen that $T_{eq} > T_d$ does not hold for $n=4$ in any
case. Thus, there is no working case for the curvaton scenario in the
positive thermal logarithmic potential. 

\section{Thermal logarithmic potential of $LLe$ and $QuQd$ flat
directions} 
Here we first derive explicitly the negative thermal logarithmic
potential arises taking $n=6$ $L_1 L_2 e_3$ direction for example. In
the end, we also show for $n=4$ $Q_1 u_2 Q_2 d_3$ direction. 
Generally, the free energy depending the gauge coupling is provided by
two loop diagrams as $V=A g_a^2(T)T^4$, where $A$ is a numerical
factor, and $a=1,2,$ and 3. Usually, $g_a(T)$ can be determined from
the renormalization group equation
\begin{equation}
    \frac{d}{dt}g_a = \frac{1}{16\pi^2}b_a g_a^3,
\end{equation}
where $t=\ln(\mu)$, and $b_a=11$, $1$, and $-3$, respectively for
$U(1)_Y$, $SU(2)_L$, and $SU(3)_C$ in MSSM. However, in the presence
of the flat direction, those fields coupled directly to it acquire the
mass of order $f \phi$, where $f$ is gauge or Yukawa coupling. Then,
the runnings of gauge couplings (values of $b_a$) change at the scale
$\mu=f\phi$ accordingly. We will write the new values by
$\tilde{b}_a$. Thus, the difference of the gauge coupling at the scale
$\mu=T<f \phi$ becomes
\begin{equation}
    \Delta g_a(T) \simeq \frac{b_a -\tilde{b}_a}{32\pi^2}
    g_a^3(T)\big|_{\phi=0} \log \left(\frac{f^2\phi^2}{T^2}\right),
\end{equation}
where $g_a(T)|_{\phi=0}$ is the coupling without the presence of the
flat direction. Since the leading term of the potential dependent on
$\phi$ is written as $V=2 g_a \Delta g_a(\phi) T^4$, the thermal
logarithmic potential is derived as
\begin{equation}
    V_{T2}(\phi) \simeq A \frac{b_a -\tilde{b}_a}{16\pi^2}
       g_a^4(T)\big|_{\phi=0} T^4 \, 
       \log \left(\frac{\phi^2}{T^2}\right).
\end{equation}

Now we have to evaluate $A$ and $b_a$ for the presence of the flat
direction. We take an explicit example of $L_1 L_2 e_3$ for
simplicity. Because of the flat direction, only $Q$, $u$, $d$, $H_u$,
and $SU(3)$ gauge fields are in thermal bath. Since the $g_3$-running
will not change, we have to see the changes of the runnings of $g_2$
and $g_1$. They are easily derived as $\tilde{b}_2=5$ and
$\tilde{b}_1=6$. The only diagrams that can contribute to the free
energy of the type $g_a^2 T^4$ are those four scalar interactions in
the D-term. Counting the number of degrees of freedom, we obtain 
$A_2 = 5/96$ and $A_1 = 7/288$. Thus,
\begin{eqnarray}
    V_{T2}(\phi) & \simeq & \left( 
      \frac{35}{4608\pi^2} R - \frac{5}{384\pi^2} \right) 
            g^4(T) \ T^4 \log \left(\frac{\phi^2}{T^2}\right), \\
    & \simeq & - f_{eff}^4 T^4 \log \left(\frac{\phi^2}{T^2}\right),
\end{eqnarray}
where $R=(g'/g)^4$, and the last line is evaluated at $\mu=T=10^{10}$
GeV (see Eq.(\ref{Ttr})), so $f_{eff} \approx 0.1$. This is the
ideal example for the scenario of the phase direction as a curvaton 
where the trap is released during IOD. 

For $Q_1 u_2 Q_2 d_3$ direction, we only show the final result:
\begin{equation}
     V_{T2}(\phi) \simeq -\frac{1}{16\pi^2\cdot 576}
     \left(\frac{427}{2}+
       \frac{9}{2}R_2 - \frac{1829}{18}R_1 \right)
        g_3^4(T) \ T^4 \log \left(\frac{\phi^2}{T^2}\right),
\end{equation}
where $R_2 = (g_2/g_3)^4$ and $R_1 = (g'/g_3)^4$. Evaluating at 
$\mu \simeq 2\times 10^{11}$ GeV, we obtain $f_{eff} \approx 0.2$,
which gives the perfect example for the scenario of the radial
direction as a curvaton when $Q<Q_c$.

\end{document}